\documentclass[referee]{sn-jnl}

\usepackage[document]{ragged2e}

\jyear{2022}%

\theoremstyle{thmstyleone}%

\theoremstyle{thmstyletwo}%

\theoremstyle{thmstylethree}%

\unnumbered%

\newcommand{\etal}{\textit{et al.~}}

\begin{document}

\title[Statistical Design for Dielectric Materials]{Identification of High-Dielectric Constant Compounds from Statistical Design}

\author[1]{\fnm{Abhijith} \sur{Gopakumar}}\email{abhijithmg@gmail.com}

\author[1]{\fnm{Koushik} \sur{Pal}}\email{koushik.pal.physics@gmail.com }

\author*[1]{\fnm{Chris} \sur{Wolverton}}\email{c-wolverton@northwestern.edu}

\affil*[1]{\orgdiv{Department of Materials Science and Engineering}, \orgname{Northwestern University}, \orgaddress{\street{2220 Campus Drive}, \city{Evanston}, \postcode{60208}, \state{IL}, \country{USA}}}

\abstract{The discovery of high-dielectric materials is crucial to increasing the efficiency of electronic devices and batteries. Here, we report three previously unexplored materials with very high dielectric constants (69 $<$ $\epsilon$ $<$ 101) and large band gaps (2.9$<$ $E_{\text{g}}$(eV) $<$ 5.5) obtained by screening materials databases using statistical optimization algorithms aided by artificial neural networks (ANN).  Two of these new dielectrics are mixed-anion compounds (Eu$_5$SiCl$_6$O$_4$ and  HoClO), and are shown to be thermodynamically stable against common semiconductors via phase-diagram analysis. We also uncovered four other materials with relatively large dielectric constants (20$<$$\epsilon$$<$40) and band gaps (2.3$<$$E_{\text{g}}$(eV)$<$2.7). While the ANN training data is obtained from Materials Project, the  search-space consists of materials from Open Quantum Materials Database (OQMD) - demonstrating a successful implementation of cross-database materials design. Overall, we report dielectric properties of 17 materials calculated using ab-initio calculations, that were selected in our design workflow. The dielectric materials with high dielectric properties predicted in this work open up further experimental research opportunities.}

\maketitle

\section{Introduction}
    Dielectric materials are among the most vital components for microelectronic device manufacturing. They are used in memory devices, capacitor based energy storage, field effect transistors, etc\cite{ortiz2009high,wang2018high, kingon2000alternative}. The dielectric constant (denoted here as $\epsilon$), more commonly referred to as the relative permittivity, is the factor by which the electric field strength decreases inside a material compared to the vacuum when it is placed near a finite electric charge. The  $\epsilon$ values of commonly used dielectric materials range between 20 and 30\cite{ortiz2009high, shevlin2005ab, delugas2007cation} - for example, Ta$_2$O$_5$ ($\epsilon$ $\sim$ 23-27, $E_{\text{g}}$=4.2 eV)\cite{ortiz2009high, iino2003organic, wang2018high, kukli1995properties} and TiO$_2$ ($\epsilon$=27, $E_{\text{g}}$=3.5 eV)\cite{ortiz2009high, ramajothi2008performance, wang2018high}. There is a high demand to find novel materials with high $\epsilon$ to increase the device performance and reliability. Typically, $\epsilon$ and $E_{\text{g}}$ are  inversely related\cite{wang2018high,lee2018high} in a compound. As a result, although several materials are reported to have an even larger $\epsilon$ values, they often have a small $E_{\text{g}}$\cite{wilk2001high, lee2018high, petretto2018high, petousis2017high}, making the dielectric vulnerable to leakage currents under exposure to large electric fields\cite{ortiz2009high,wang2018high}. Therefore, compounds with high $\epsilon$ and large band gaps are preferred while designing charge storage applications and microelectronic devices.

    One of the methods to find high-$\epsilon$ compounds is to calculate the dielectric constants and band gaps of a large number of compounds that are available in large materials databases such as the Open Quantum Materials Database (OQMD)\cite{saal2013materials, kirklin2015open}, Materials Project (MP)\cite{Jain2013}, etc using ab-initio methods such as density functional theory (DFT). However, since the accurate calculation of dielectric properties using density functional perturbation theory\cite{giannozzi2005density} (DFPT) is computationally very expensive, it would be practically unfeasible to estimate the dielectric constants of tens of thousands of materials available in those database using high-throughput methods. In this work, we employ an advanced  screening strategy to identify compounds with better dielectric properties. Thus, the goal of this work is to find dielectric materials with large values for both $\epsilon$ and $E_{\text{g}}$ by screening materials databases but at the expense of conducting as few DFPT calculations as possible. To accomplish this task, we have employed a materials design strategy comprised of statistical optimization models and  DFPT calculations on a small set of compounds. While our training set consists of a small amount of data (dielectric constants) from the MP, the search-space contains a vast set of compounds available in the OQMD.

    Several online data repositories exist today that are dedicated to hosting large sets of open-sourced inorganic crystal structure data generated from high throughput (HT) DFT calculations such as the MP\cite{Jain2013}, OQMD\cite{saal2013materials, kirklin2015open}, and AFLOWLib\cite{curtarolo2012aflowlib} among others\cite{draxl2019nomad, choudhary2020high}. The design and discovery of novel materials using statistical modelling has become an active research area \cite{pyzer2015learning,saal2020machine,park2020developing} in recent times, largely attributed to the availability of such HT datasets. Recently, multiple studies have reported HT-generation of dielectric data and subsequent analysis\cite{lee2018high,umeda2019prediction,qu2020high}. For example, Morita \etal reported\cite{morita2020modeling} machine learning modeling of data from MP \cite{petousis2017high, petretto2018high, Jain2013} to assess the reliability of the theoretical models currently available to describe the dielectric properties of crystals.
    
    In this work, we use the MP-dataset of 1864 dielectric tensors\cite{petretto2018high,petousis2017high} to train statistical models and subsequently identify dielectrics from the set of stable materials in the OQMD. Thus the MP-data forms the training-data and the set of materials from OQMD forms the search-space for the materials design. This work is a successful demonstration of the scenario where the data obtained from the multiple sources can be utilized to discover new compounds. The negligible difference found between the representation vectors, which are also called as feature vectors in machine learning, generated for equivalent materials in MP and OQMD made the cross-database design possible in this work. Overall, we conducted three design cycles which required us to perform dielectric calculations for just 17 materials using DFPT. We report the dielectric constant values of all the 17 materials among which three of them (HoClO, Eu$_5$SiCl$_6$O$_4$, and Tl$_3$PbBr$_5$) have very large $\epsilon$ (69 $ < \epsilon < $101) and $E_{\text{g}}$  (2.9 eV $ < E_{\text{g}} < $5.5 eV) values making them part of the Pareto-front of the known data, and four other materials (Sr$_2$LuBiO$_6$, Bi$_5$IO$_7$, Bi$_3$ClO$_4$, and Bi$_3$BrO$_4$) have moderately large $\epsilon$  (20 $ < \epsilon < $40) and $E_{\text{g}}$ (2.3 eV $ < E_{\text{g}} < $2.7 eV) values.

\section{Results}
\subsection{Materials design strategy}
\begin{figure*}[!ht]
    \centering
    \includegraphics[width=1.0\linewidth]{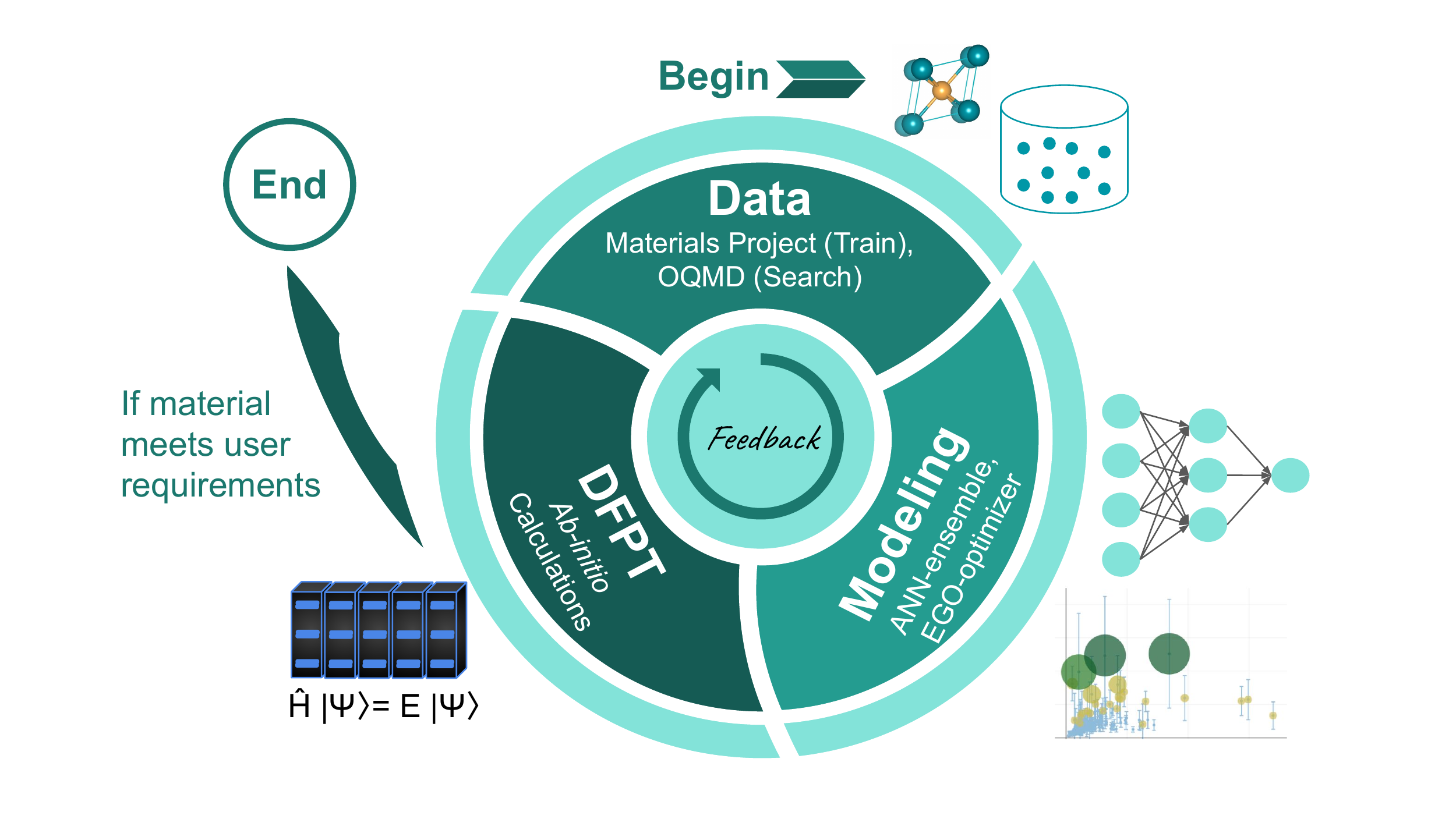}
    \caption{\textbf{Materials Design workflow used in this work.} The three parts of the design workflow shown here together completes a single design cycles. The newly computed DFPT results from a design cycle is fed back into the training-data for the next design cycle. Since the dielectric tensor of a crystal is of shape 3$\times$3, optimizing all nine components of  the dielectric tensor leads to a nine-objective optimization problem.  Thus, we specifically optimize the largest value of dielectric constant ($\epsilon$) among all crystallographic directions, also referred as the target-property hereinafter. This scalar value is quantified as the largest eigenvalue of the total dielectric tensor. The training-data in this work consists of nearly 2000 compounds from the MP for which $\epsilon$ and $E_{\text{g}}$ were known and the training-data came from the OQMD, for which only $E_{\text{g}}$ were known at the beginning of this work.}
    \label{fig:designflow}
\end{figure*}

    Our objective is to find large band gap materials with optimal dielectric constants. Since, the dielectric tensor of a compound has nine components, the optimization of all nine components leads to a nine-objective optimization problem which is difficult to solve with training-data of size $\sim$ 2000. Thus, we specifically optimize the largest eigenvalue of the dielectric tensor, referred to from here onward as $\epsilon$, via statistical modelling through the materials design workflow, as depicted in Figure \ref{fig:designflow}. The workflow is similar to the strategies that have been previously reported in literature\cite{balachandran2016adaptive, gopakumar2018multi}, where each design cycle consists of three steps - data processing, statistical modeling, and ab-initio DFPT calculations. The largest eigenvalue of the total dielectric tensor is chosen as the property to be optimized because that is the highest possible dielectric behavior from a single crystal when it is aligned perfectly along the corresponding direction between two metallic plates. The total dielectric tensor is calculated as the sum of ionic and electronic dielectric tensors. The good agreement between dielectric tensor eigenvalues obtained from MP's DFPT high throughput framework and experimentally measured dielectric constant values was reported by Petousis et al.\cite{petousis2016benchmarking}. We preferred the largest eigenvalue over the average of eigenvalues because the latter value may severely underestimate the highest possible dielectric behavior from a single-crystal (Supplementary Figure 1), even though it is a popular choice to estimate the polycrystalline dielectric constant\cite{petousis2016benchmarking,petousis2017high}. The new data produced from DFPT calculations at the end of each cycle is fed into the next design cycle. In the first step, we collected the relevant data from the MP database (training-data) and OQMD (search-space). All materials in the training-data have a known value for $\epsilon$ and $E_{\text{g}}$, while the materials in search-space have known values of $E_{\text{g}}$ but their $\epsilon$ values are unknown. In the second step, Modeling, we created an ensemble of artificial neural network (ANN)\cite{jain1996artificial} models, fit on the training-data, which learn to predict the $\epsilon$ value of materials when their crystal structures and $E_{\text{g}}$ values are known. Using this ANN-ensemble, we predicted the $\epsilon$ of each material in the search-space. Since the prediction was done from an ensemble, the results were a distribution of $\epsilon$ values for each material, contrary to the usage of a single ANN-model where a single prediction-value is obtained.  The trained ANN ensemble was used to predict the $\epsilon$-distributions of 11,102 stable non-metallic materials in the search-space, obtained from the OQMD. 
    
    Further, the predicted distribution of $\epsilon$ was input into the Efficient Global Optimization (EGO)\cite{balachandran2016adaptive} algorithm. EGO takes into account of the distribution's mean and standard-deviation to rank the materials in search-space based on their potential to increase the chances of finding high-$\epsilon$ materials in this workflow within as few design cycles as possible. In this work, the optimization in dielectrics refers to the identification of dielectrics with large $\epsilon$ values. The reason for employing an EGO-algorithm to explore the search-space is to account for the uncertainty in ANN model predictions when the available training-data may not have sampled the material-space uniformly. The advantages of EGO-based optimization in materials design was first reported and benchmarked by Balachandran \etal \cite{balachandran2016adaptive, balachandran2017learning,balachandran2018predictions}. In this work, we used the EGO algorithm to select the best candidates that are either predicted to have high $\epsilon$ value or have a large uncertainty in their ANN-ensemble predictions. Materials that belong to the latter category are from the regions of materials yet to be sampled by the training-data. The DFPT characterization of such materials is expected to increase the reliability of ANN-ensemble predictions after each design cycle and eventually lead to better optimization of dielectrics during the course of this work.
    
    The metric that is used to rank the materials is called Expected Improvement, or $E(I)$. More details on how the $E(I)$ is calculated, is provided in the Methods section. A few (5-6) materials were selected in this step with highest values of $E(I)$ and carried onto the next step - DFPT calculations. In this final step, the dielectric tensor of the selected materials were calculated using  DFPT calculations. If DFPT results show that any of the materials have a high value of $E_{\text{g}}$ and $\epsilon$, we stop the design workflow at that point. Otherwise, a new design cycle is started after transferring the newly computed $\epsilon$ values and the corresponding materials to the training-data from the search-space. With an increased size of training-data, the ANN-ensemble is expected to have less uncertainty in $\epsilon$ predictions in the new design cycle. The design cycle was repeated with feedback three times in total in this work until three materials with very large values for $E_{\text{g}}$ and $\epsilon$ were found.

\subsection{Data}

    A dataset containing information about crystal structures, chemical compositions, band-gap energy values and dielectric-tensors of 1864 stable materials was obtained from the MP\cite{petousis2017high,petretto2018high,Jain2013} data repository. This dataset was used to generate the training-data. The target-property, $\epsilon$, was obtained for each material in this database from its calculated dielectric-tensor. Another dataset consisting of 11,102 stable, non-metallic materials containing information about crystal structures, chemical compositions, band-gap energy values was obtained from OQMD\cite{saal2013materials, kirklin2015open}. This OQMD dataset was used to generate the search-space in which the search to find dielectrics was conducted. The dielectric tensor data of all crystals included in the search-space were unknown at the beginning of this work.

    The materials need to be represented as vectors of uniform length in order to be input into a statistical model. We generated the material representations using the Magpie \cite{ward2016general} crystal property generator tool. Magpie generates a set of physical features (such as the mean electronegativity of constituent atoms, average coordination number inside unit cell, etc) from a given chemical composition and crystal structure. Within Magpie, the crystal's structure-related features are generated by building Voronoi tessellations inside the crystal and finding nearest neighbors of each individual atom\cite{ward2017including}. Magpie generated 271 input features that include 145 composition-based, and 126 structure-based features to represent each material. In addition to these, the material's DFT $E_{\text{g}}$ value was also added as an extra feature to the representation-vector since it is already known for all materials in both MP and OQMD datasets. Addition of $E_{\text{g}}$ increased the size of representation-vector to 272, which was generated for each material in training-data and search-space. The input feature-vector size was further reduced to 100 using the widely-used feature reduction techniques such as principal component analysis and model-based selection, implemented in the Scikit-learn python library\cite{scikit-learn}. The set of material representation-vectors of training-data and the search-space, in addition to the target values associated with the training-data, completes the first step of materials design as depicted in Figure \ref{fig:designflow}. The size of the training-dataset increases after each design cycle as a result of conducting DFPT calculations on new materials from the search-space.

\begin{figure*}[!ht]
    \centering
    \includegraphics[width=0.9\linewidth]{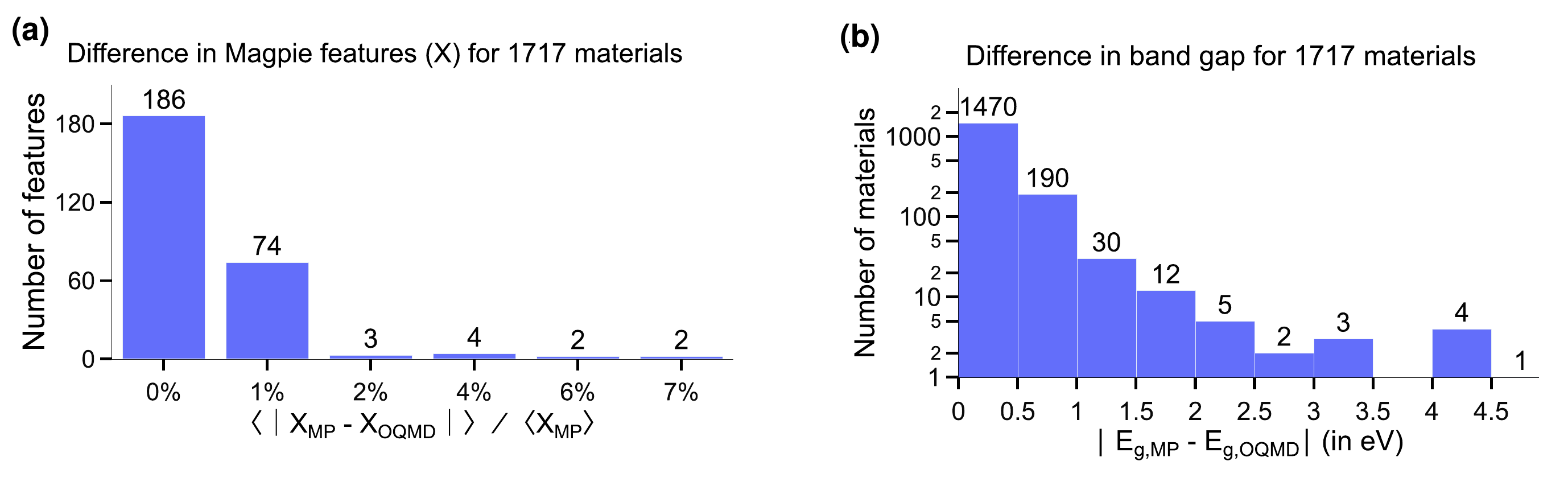}
    \caption{\textbf{Comparison of material representation vectors between the OQMD and MP structures.} Difference in material representation vectors of the structures obtained from OQMD and MP for 1717 materials. \textbf{a)} Mean absolute difference in the Magpie-generated representational feature vectors on structures obtained from the MP and OQMD for 1717 materials in the training-data. Crystal structures of all 1717 materials were first obtained from MP as a part of generating training-data, and further cross referenced to find their equivalent structures in OQMD based on their ICSD Collection Codes. ICSD Collection Codes were not available for the rest of the 143 materials in the MP training-data. \textbf{b)} Difference in band gap values of 1717 materials from training-data that have a corresponding structure entry in MP and OQMD, which are cross-referenced based on their ICSD Collection Codes.}
    \label{fig:cross_db_reasoning}
\end{figure*}

    Statistical modelling utilizing data from multiple computational material databases is prone to error arising from the differences in the DFT parameters used at each database's high-throughput calculation strategy. Here, we have investigated the difference in Magpie-generated features for equivalent materials in OQMD and MP, cross referenced based on their associated Inorganic Crystal Structure Database\cite{belsky2002new} (ICSD) Collection Codes. In total, 1717 out of 1864 materials in training data had an ICSD Collection Code associated with them. The crystal structures from OQMD corresponding to all the 1717 ICSD materials were obtained, and their Magpie-generated features were compared against that of the structures obtained from MP as a part of the training data. The results, as plotted in Figure \ref{fig:cross_db_reasoning}a, shows negligible ($\leq 2\%$) relative difference in 263 out of a total of 271 Magpie features, while the other eight features have low relative differences ($\leq 7\%$). All 145 composition-based features are computed to be identical across the databases, as expected. The finite difference in some of the structure-based features originate because of the difference in accuracy of crystal structural minimization across databases.
    Band gap, which joins the Magpie features to form the final material representation-vector, was also compared between OQMD and MP for the 1717 equivalent materials, as shown in Figure \ref{fig:cross_db_reasoning}b. Band gap values showed a mean and median absolute deviation of 0.1 eV and 0.0 eV respectively, pointing toward a negligible difference between the calculations of band gap for materials included in the training data across OQMD and MP. Overall, the materials representation-vector considered in this design is generated in a cross-comparable manner across OQMD and MP structures with very low errors.
    
    The $\epsilon$ values in the training-data obtained from MP are predominantly concentrated in the range of 0 to 25, making it difficult to model the data reliably for materials with large $\epsilon$ due to a possible bias toward smaller values. Less than 5\% of the materials in the training-data have $\epsilon$ $>$ 50. The median of $\epsilon$ values in MP-dataset is 12.2 while the mean and standard deviation are 20.2 and 42.8 respectively. The distribution of $\epsilon$ in training-data is shown in Supplementary Figure 2. The large spread of $\epsilon$ values is decreased upon a log-scale transformation, as shown in Figure \ref{fig:model_ei}a. A smaller spread of target values helps stabilize the machine learning model during the training by reducing the probability of excessive changes in internal parameters, such as the weights in an ANN. 
    We also analyzed the correlation between $\epsilon$ and $E_{\text{g}}$ values for the materials in the training data, and it is given in Supplementary Figure 3.

    The original dataset downloaded from MP listed BeO (MP ID: mp-1794) as having large \textit{ab-initio} computed values for $\epsilon$(=312) and $E_{\text{g}}$(=8.2 eV). This large value of $\epsilon$ is possibly caused by the improper relaxation of the primitive cell of BeO in MP that leads to a large volume change. Hence, the succeeding calculations on this compound such as DFPT may be incorrect. We conducted a separate DFT cell-relaxation and DFPT calculation for BeO using VASP starting with the MP's initial structure and find that the computed $\epsilon$ value for the correctly relaxed structure is 4 - well in agreement with the previously reported values in literature\cite{groh2009first}. This compound was removed from the training-data before proceeding further. We looked up other materials in training-data with very high $\epsilon$ and smaller $E_{\text{g}}$ individually and confirmed that they did not have a large cell-volume change upon relaxation in MP.

\subsection{Statistical modeling}
    The predictions from trained machine learning models, such as ANNs, are often prone to errors arising from the insufficient sampling of material space by training-data. We needed to quantify the uncertainty associated with the $\epsilon$ value predictions even though the available ANN algorithms explicitly do not provide that value from a single ANN model. So we created an ensemble of ANNs, each of which was trained on a randomly chosen subset of the training-data, and has different architectures and internal parameters. An ANN-ensemble containing 2000 independent ANN models was created and trained at each design cycle. Each ANN in the ensemble predicted a single $\epsilon$ value upon inputting a material-representation vector, resulting in a distribution of 2000 predicted $\epsilon$ values for each material in the search-space. The standard deviation of each of the predicted $\epsilon$-distribution was defined as the uncertainty of ANN modeling for the corresponding material. 

\begin{figure*}[!ht]
    \centering
    \includegraphics[width=0.9\linewidth]{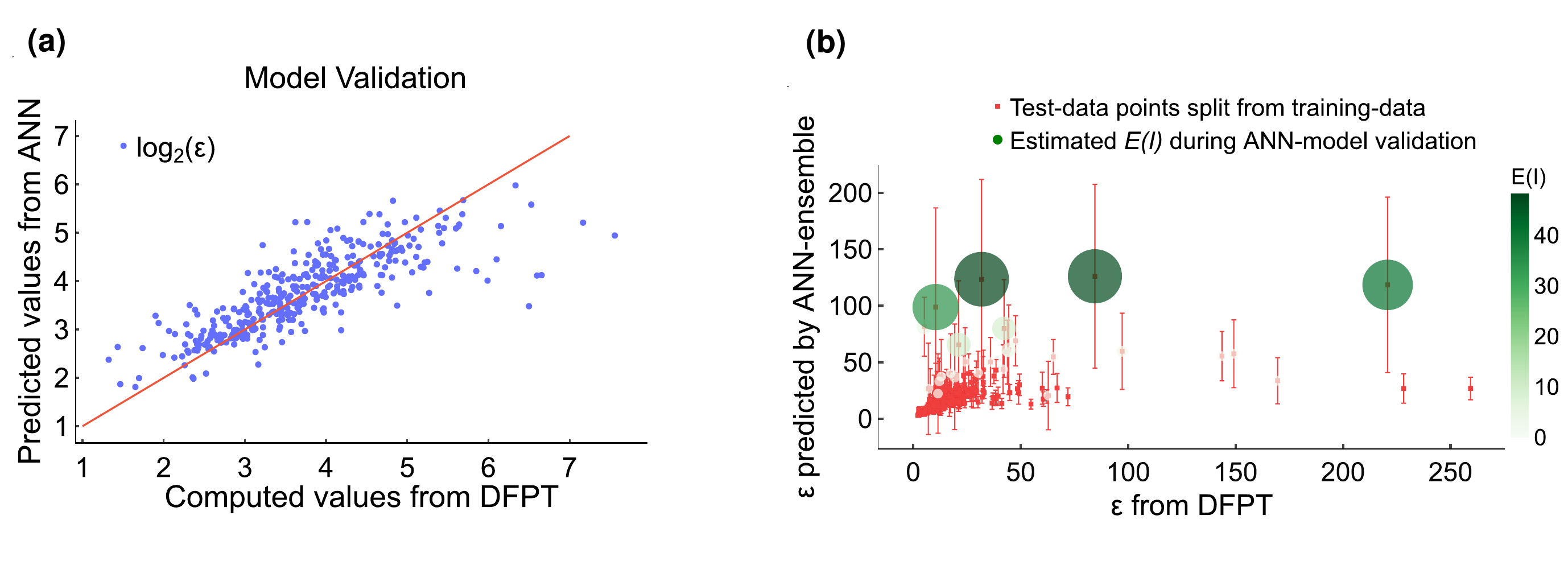}
    \caption{\textbf{Results from statistical modeling.} \textbf{(a)} ANN model validation on a test set of 373 materials split from the training-data. We used an ensemble of ANNs to predict a distribution of values for each material. This particular model-fit plot is taken from a single ANN-model that was part of the ensemble in design cycle 2. The 373 materials plotted here were not seen by this particular ANN model at any stage during the training. These predictions are made only for this particular ANN model to show its learning capabilities, and it is not part of the design workflow that we created. In the design workflow, each ANN model in the ensemble is exposed only to a unique subset of the full MP training data, excluding 373 randomly chosen materials. Further, in the design workflow, this trained ANN model is used to predict the dielectric values of only the search-space materials from OQMD, not the 373 unseen materials from MP dataset. The model was trained to predict $\log_2(\epsilon)$ because the $\epsilon$ values were highly non-uniform in the training-data with most of the values were below 25, making some of the very large values as outliers. A log-scale transformation of $\epsilon$ reduced the numerical difference between the largest $\epsilon$ value and the median, making the former less of an outlier in ANN modeling. The model fit shown in this plot has an $R^2$ score of 70\%, and a Spearman's rank correlation of 85\%. \textbf{(b)} This plot shows the predicted $\epsilon$-distributions and corresponding $E(I)$ values on the same test dataset consisting of 373 materials split from the training-data. The error bars represent the standard deviation in ANN-ensemble predictions which is quantified as the uncertainty of ANN modeling. For a clearer perspective, the radius and color of the circles represent the same quantity - the Expected Improvement,$E(I)$, value calculated using EGO algorithm. A point without an outer circle around it represents a material with a negligible ($<10^{-3}$) value for $E(I)$. In this figure, only 25 materials have an $E(I)$ value that is greater than $10^{-3}$.}
        \label{fig:model_ei}
\end{figure*}

    Further, a statistical single-objective optimization algorithm, called EGO \cite{xue2016accelerated,balachandran2016adaptive,jones1998efficient,solomou2018multi,talapatra2018autonomous}, was used in this work to evaluate the $\epsilon$-distribution and quantify a measure of probable optimization associated with each material in the search-space. EGO is not a method to model the data and predict $\epsilon$. Instead, EGO is an algorithm to select the best candidates from a given search-space, based on their $\epsilon$-distributions predicted by the ANN-ensemble, in order to discover as many high-$\epsilon$ materials from as few design cycles as possible. Here, the desired optimization is the maximization of $\epsilon$ among all the materials in search-space. The quantified measure of predicted optimization in EGO is called expected improvement, denoted as $E(I)$. Conceptually, the $E(I)$ of a material in search-space is the quantified probability with which a DFPT calculation of $\epsilon$ for that material will lead to a identification of high-$\epsilon$ material in the design workflow within as few design cycles as possible. 
    Figure \ref{fig:model_ei}a shows the results from an ANN model validation as a part of model training during the second design cycle. The values of $E(I)$ computed for the same validation data split from the training-data, is shown in Figure \ref{fig:model_ei}b.
    A simplified illustration of $E(I)$ with the help an example is given below.

\subsection{Example Illustration of \textit{E(I)}}
    Suppose the predicted $\epsilon$-distribution belonging to a material $M_1$ in the search-space has a large standard deviation. Then it is highly probable that the material $M_1$ belongs to a part of material representation-vector space which was not sampled very well in the training set. Computing the $\epsilon$ of $M_1$ using DFPT and feeding back that information to the training-data will lead to better ANN modeling in the subsequent design cycles. Thus,  $M_1$ will have a large value of $E(I)$. Now consider another material $M_2$ in search-space with a large mean and a small standard deviation for its predicted $\epsilon$-distribution. The material $M_2$ belongs to a part of the material representation-vector space that was sufficiently sampled by the training-data. So it is highly probable that $M_2$ will turn out to be a high-$\epsilon$ material upon DFPT calculations. Because of that, $M_2$ will also have a large value of $E(I)$. 

\begin{figure*}
    \centering
    \includegraphics[width=0.9\linewidth]{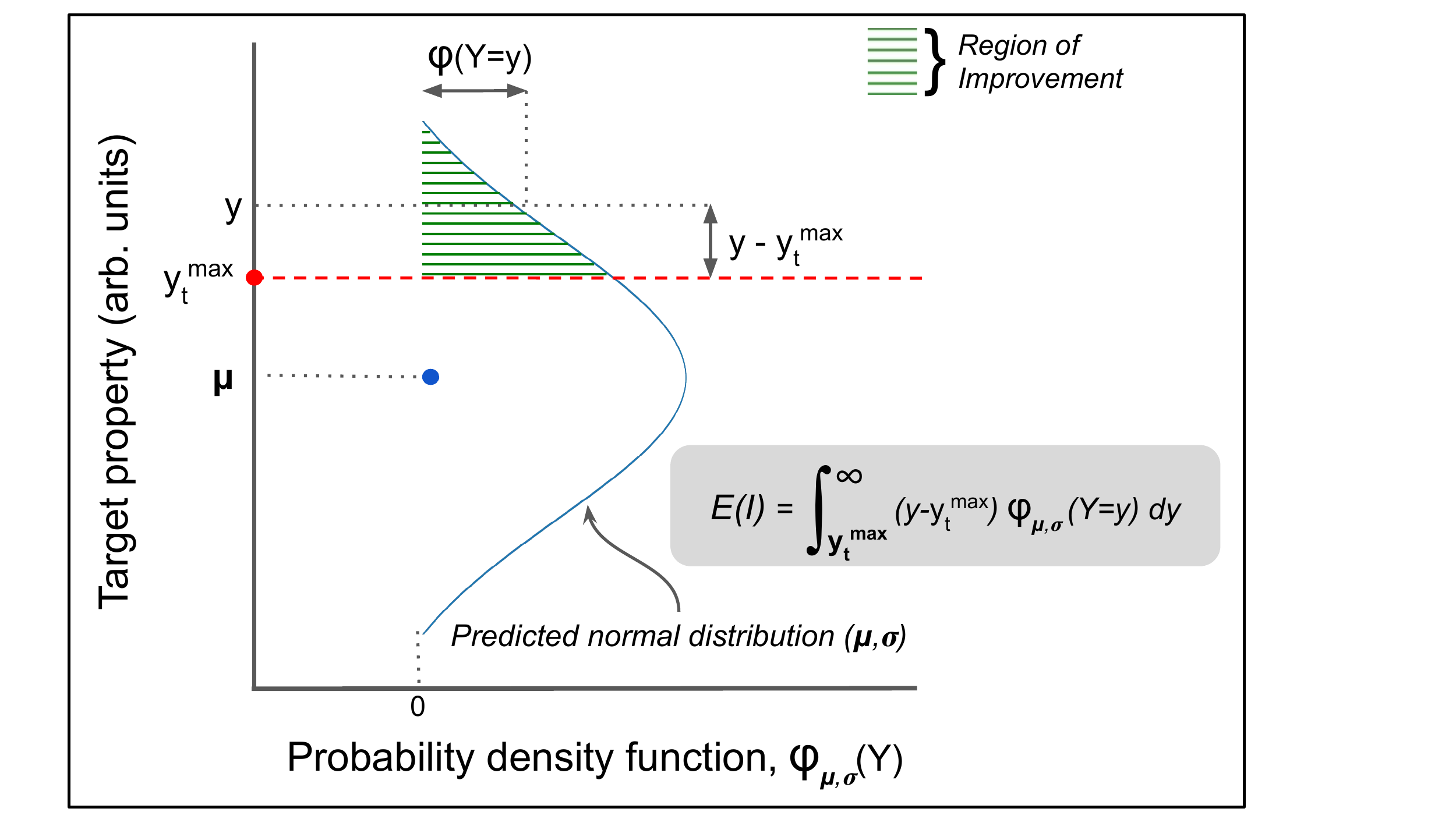}
    \caption{\textbf{The optimization algorithm.} The value $y_t^{max}$ represents the currently available highest value of $\epsilon$ among all materials in the training-data. $\mu$ and $\sigma$ represents the mean and standard deviation of the ANN-ensemble predicted distribution of $\epsilon$ for a material (blue dot) in the search-space. Within the predicted distribution, which is assumed as a normalized Gaussian function here, the region above $y_t^{max}$ represents the region of improvement - as shown in green. If the ab-initio DFPT calculation determines that the material's $\epsilon$ value exists within the green-shaded region, it will be considered as an improvement over the current best value $y_t^{max}$ in training-data.}
    \label{fig:ei_explanation}
\end{figure*}

    In EGO, the calculation of $E(I)$ for a general optimization problem proceeds as follows (also shown in Figure \ref{fig:ei_explanation}):\\
    Let $Y$ be the target property to be maximized and $\varphi(Y)$ be the predicted distribution of $Y$ for a given search-space material. The value, $\varphi(Y=y)$ is the probability when the  value of $Y$ is $y$. The largest value of target property in the training-data is denoted as $y_t^{max}$. The EGO algorithm, as formulated by Jones et al.\cite{jones1998efficient}, computes the expected improvement, $E(I)$, as
    \begin{equation}
        E(I) = \int_{y_t^{max}}^{\infty} (y-y_t^{max}) \; \varphi(Y=y) \; dy
    \end{equation}
    As mentioned in Balachandran \etal \cite{balachandran2016adaptive}, if the predicted distribution is approximated as a normal (i.e., Gaussian) distribution with a mean $\mu$ and a standard deviation $\sigma$, the above equation can be re-written as,
    \begin{equation}
        E(I) = \sigma[\phi(z) + z \Phi(z)]
    \end{equation}
    where, $z = \frac{\mu-y_t^{max}}{\sigma}$, $\phi$ is the probability density function, and $\Phi$ is the cumulative distribution function\cite{jones1998efficient} of the normal distribution, $\varphi(Y)$.

    For dielectric design, $Y$ is the dielectric constant ($\epsilon$) of a candidate material, and $y_t^{max}$ is the highest value of $\epsilon$ in the training-data obtained from DFPT calculations. In the MP-dataset, the largest $\epsilon$ value is for TiO$_2$ with $\epsilon$=988 and $E_{\text{g}}$=1.8 eV. But our goal in this work is to find materials with large $\epsilon$'s, not necessarily higher than 988 as long as the $E_{\text{g}}$'s are greater than 1.8 eV. Thus the $y_t^{max}$ in this work was set at 100.0 for all design cycles, instead of setting it at 988.0, to consider the search-space materials whose $\epsilon$ values are predicted to be sufficiently high. The $\varphi(Y)$ is approximated to be a normal distribution with the same mean, $\mu$, and standard deviation, $\sigma$, as that of the original $\epsilon$-distribution predicted by the ANN-ensemble for each search-space material.

\subsection{Design cycles with feedback}
    The $\epsilon$ values of a few materials selected from the statistical modeling are computed from DFPT calculations, as shown in the final segment of a design cycle in Figure \ref{fig:designflow}. The results from the DFPT calculations are used to determine whether to conduct any further design cycles. In this work, we conducted the design cycles until at least one high-$\epsilon$ dielectric with a large $E_{\text{g}}$ is identified. When no such materials are found during a design cycle, all the selected materials along with their newly DFPT-estimated $\epsilon$ values are transferred from search-space to training-data, resulting in a feedback of information prior to the beginning of next design cycle. The feedback is one of the most crucial parts of our material design workflow, because it results in a better sampling of material representation-vector space by training-data and thus, more reliable ANN model predictions during the next design cycle. The advantage of the feedback mechanism is prominent during the quantification of uncertainty which is used directly by the EGO algorithm to identify the best candidates for the next set of DFPT calculations. After the end of a design cycle, the uncertainty on predicting the $\epsilon$ values is decreased for the set of materials which are similar to the materials whose $\epsilon$ values were calculated using DFPT in the given cycle. 
    
    In addition to the feedback mechanism, another factor that influenced the candidate selection in the design workflow is the minimum cutoff imposed on the band gap values of materials when they are included on the search-space. The reason for implementing a cutoff is to externally introduce a character of multi-objective optimization in this work. Without explicitly setting a minimum band gap limit, the candidate selection process that is dictated  by the EGO algorithm tries to optimize only a single objective, which is  the $\epsilon$ value. We conducted three design cycles  sequentially with feedback of the newly calculated data into training-data after each cycle. In the first design cycle, we set no band gap minimum cutoffs to allow the full exploration of the search-space that consists of 11,102 non-metals from OQMD. In the second design cycle, a minimum cutoff of 2.25 eV was set, leaving 6191 materials in the search-space. In the final cycle, the minimum cutoff was increased to 5 eV to limit the candidate selection only to the materials with very high-$E_{\text{g}}$. Hence, the search-space size in the final cycle was reduced to 1046 materials. The workflow that we adopted in this work deviates from the ideal situation where a dedicated multi-objective optimization statistical algorithm will be used to find a material with high $\epsilon$ and large $E_{\text{g}}$ values. Since the band gap values are already available for all materials in the search-space, the best approach here was to implement a statistical optimization algorithm to quickly find high-$\epsilon$ materials while the preference for large band gap values is achieved by manually setting a minimum cutoff. This work stands as an example for the modifications required to practically implement the statistical algorithms that are often benchmarked on idealistic scenarios.

\subsection{New dielectric materials}
\begin{figure*}[tp]
    \centering
    \includegraphics[width=0.9\linewidth]{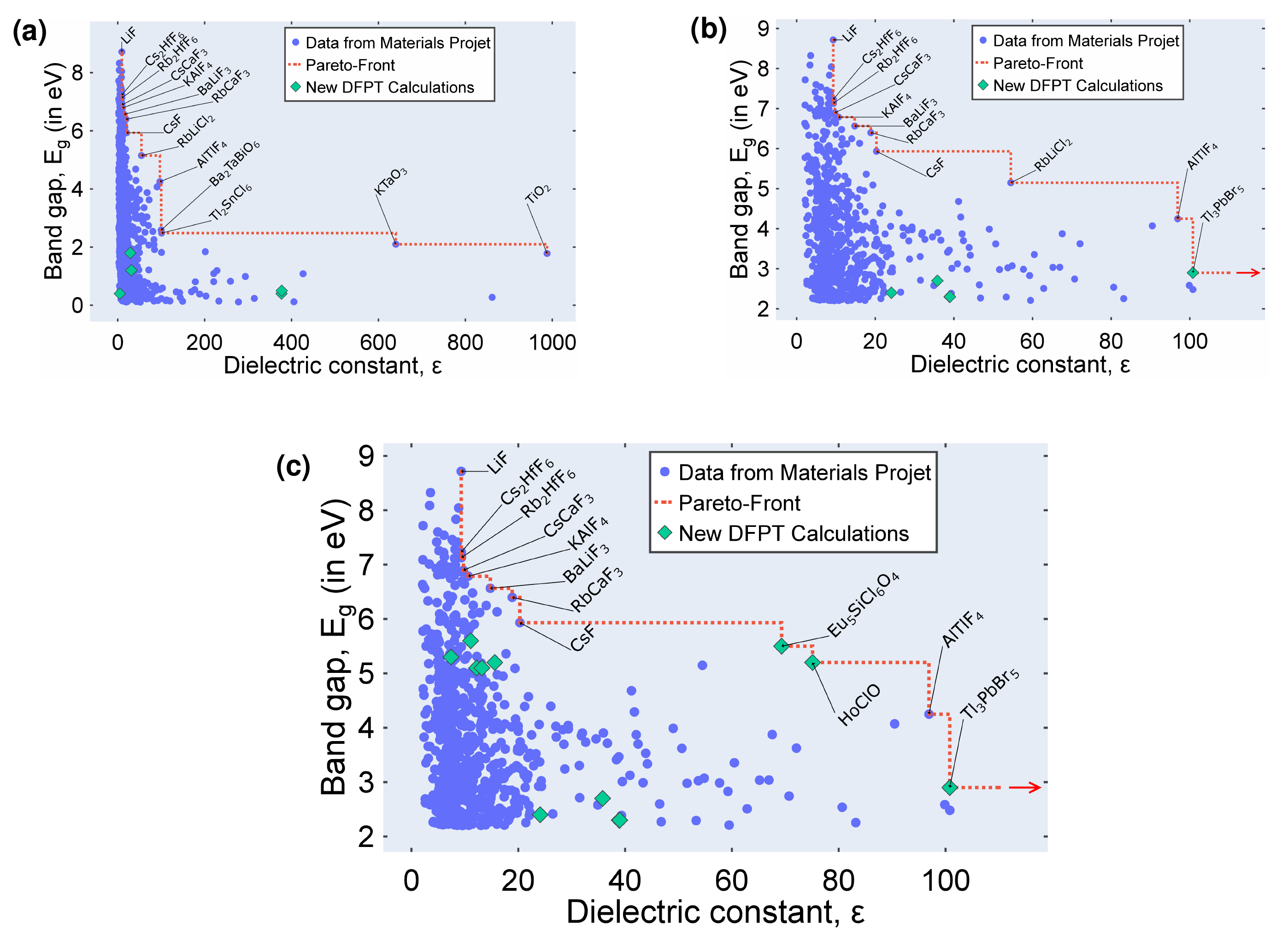}
    \caption{\textbf{Evolution of the Pareto-front with design cycles.} The $\epsilon$ and $E_{\text{g}}$ values of the training-data and newly characterized dielectric materials are plotted for \textbf{(a)} design cycle 1, \textbf{(b)} design cycle 2 and \textbf{(c)} design cycle 3. All the data shown in these plots originated from DFPT calculations. Plot \textbf{(a)} shows the original Pareto-front of the dataset from materials project (MP), because none of the materials measured in cycle 1 became part of the Pareto-front - predominantly owing to their low band gap values. Assigning no restrictions on the band gap of search-space materials during the first cycle directed the design algorithm to pick two materials without any preference for large band gaps. The numerical values of the materials in Pareto-front of MP-dataset is given in Table \ref{tab:fullMPpf}. In both \textbf{(b)} and \textbf{(c)}, only the materials with $E_{\text{g}}$ values greater than 2.0 eV are plotted to highlight the area where some of the newly discovered dielectrics in their corresponding cycles joined the Pareto-front. Due to this cropping, two materials from MP-dataset which are actually in the Pareto-front in plots \textbf{(b)} and \textbf{(c)} with very high $\epsilon$ values - tetragonal TiO$_2$ ($\epsilon$=988, $E_{\text{g}}$=1.8 eV) and cubic KTaO$_3$ ($\epsilon$=640, $E_{\text{g}}$=2.1 eV), are not shown here.}
    \label{fig:pf_all_cycles}
\end{figure*}

    The materials that are part of Pareto-front of MP-data is listed in Table \ref{tab:fullMPpf}, while the Pareto-front of training-data at each design cycle is plotted in Figure \ref{fig:pf_all_cycles}. Since the maximization of $\epsilon$ and $E_{\text{g}}$ values are considered as optimal in this study, each material in the Pareto-front has a higher value of either $\epsilon$ or $E_{\text{g}}$ than any other material in the corresponding training-data. Therefore, the modification of the training-data's Pareto-front by any of the newly calculated dielectric constants after each design cycle may indicate the identification of suitable, high-dielectric materials.
    
    During the first design cycle, the EGO algorithm picked out the five most promising candidates with largest $E(I)$ values in the search-space. The $\epsilon$ values of these five selected materials were calculated using DFPT. Two materials among them turned out have very high $\epsilon$ values ($\sim$ 370) but very low $E_{\text{g}}$ ($\sim$ 0.5 eV). The low $E_{\text{g}}$ values are not unexpected since the EGO algorithm implemented in this work aims to maximize only the $\epsilon$ values. None of the materials selected in this cycle modified the Pareto-front of MP dataset, as shown in Figure \ref{fig:pf_all_cycles}a. The $\epsilon$ values of these five materials were appended to the training-data prior to starting the next design cycle. 

    Five materials were selected in the second cycle and their dielectric constants were calculated. Our calculations predict a large dielectric constant for one of the five new materials - tetragonal Tl$_3$PbBr$_5$ ($\epsilon$=101, $E_{\text{g}}$=2.9 eV). Tl$_3$PbBr$_5$ joined the Pareto-front, as shown in Figure \ref{fig:pf_all_cycles}b. Three other new materials - Bi$_5$IO$_7$ ($\epsilon$=36, $E_{\text{g}}$=2.7 eV), Bi$_3$ClO$_4$ ($\epsilon$=39, $E_{\text{g}}$=2.3 eV), and Bi$_3$BrO$_4$ ($\epsilon$=39, $E_{\text{g}}$=2.3 eV), have moderately large $\epsilon$ values, even though they did not improve the existing Pareto-front. All the five new materials were appended into the training-data before proceeding to begin the third design cycle.
    
    During the third and final design cycle consisting of only materials with very large $E_{\text{g}}$ in search-space, seven new candidate materials were selected to do DFPT calculations. Two among them - Eu$_5$SiCl$_6$O$_4$ ($\epsilon$=69, $E_{\text{g}}$=5.5 eV) and HoClO ($\epsilon$=75, $E_{\text{g}}$=5.2 eV) joined the Pareto-front due to their large $\epsilon$ and $E_{\text{g}}$ values, as shown in Figure \ref{fig:pf_all_cycles}c. In total, three new dielectric materials in the Pareto-front were discovered after three design cycles and 17 new DFPT calculations were performed in the entire workflow. No further design cycles were conducted since we have already identified multiple compounds with high $\epsilon$ and $E_{\text{g}}$, which remained unexplored experimentally.

    The $\epsilon$ values of all 17 materials which were obtained in this work are given in Table \ref{tab:all_new_meas}. The $\epsilon$ and $E_{\text{g}}$ of all materials belonging to the Pareto-front of MP-dataset is listed in Table \ref{tab:fullMPpf} for comparison. Among all the newly discovered dielectrics with large $\epsilon$ values, tetragonal HoClO and monoclinic Eu$_5$SiCl$_6$O$_4$ stand out because of their very large DFT-calculated band gap energies (5.2 eV and 5.5 eV respectively). These two rare earth oxychlorides are reported to have been experimentally synthesized \cite{templeton1953crystal, holsa2002stability, basiev2005hydration, jacobsen1994synthesis} but their dielectric properties remained unstudied to the extent of our knowledge. 
    Both of these compounds are mixed-anionic inorganic compounds - a class of emerging functional materials\cite{kageyama2018expanding}. Interestingly, the monoclinic Eu$_5$SiCl$_6$O$_4$ has 32 atoms in its primitive unit cell which often exceeds the maximum cutoff on number of atomic sites in high throughput studies involving computationally expensive material properties\cite{petretto2018high, choudhary2020high}. 
    
   Thermodynamic stability of a dielectric when in contact with Si or other semiconductors is an important requirement for it to be used in electronic applications. Several of the high-$\epsilon$ dielectrics identified in the published literature were shown to be unstable while forming an interface with Si in subsequent experimental studies conducted at or above the room temperature. The formation of SiO$_x$ and other undesired metal oxides were reported at the interface between Si and the popular high-$\epsilon$ dielectrics such as Ta$_2$O$_3$\cite{atanassova1998x, schlom2002thermodynamic, alers1998intermixing}, TiO$_2$\cite{perego2008energy, mccurdy2004investigation}, BaTiO$_3$\cite{george2013preferentially} and SrTiO$_3$\cite{hu2003interface, goncharova2006interface}. The thermodynamic stability between two compounds can be assessed from the phase diagram involving those compounds. In this work, the phase diagram is constructed by computing the convex hull\cite{barber1996quickhull} of formation energies of all the materials that belong to a given phase space spanned by their constituent elements. Each of the compounds that form the convex hull not only has the lowest formation energy at its composition but also has lower energy than any linear combination of other materials in that phase space. The difference between the formation energy of a compound and energy at the convex hull for the same composition is called as hull-distance ($E_{\text{hd}}$). By definition, each material that is on the convex hull has a hull distance  of zero (i.e., $E_{\text{hd}}$ = 0) and is considered to be stable. On the other hand, every material that falls above the convex hull is considered as metastable (0 $<$ $E_{\text{hd}}$ $\leq$ 50 meV per atom) or unstable ( $E_{\text{hd}}$ $>$ 50 meV per atom) depending on the magnitude of $E_{\text{hd}}$ according to  the heuristic conventions adopted in literature \cite{sun2016thermodynamic,balachandran2018predictions, wu2013first, zakutayev2013theoretical, pal2021accelerated}. The presence of a tie line between two compounds in a convex hull phase diagram indicate that they are thermodynamically stable phases when in contact with each other. Our thermodynamic stability analysis on Ta$_2$O$_3$, TiO$_2$, BaTiO$_3$ and SrTiO$_3$ in OQMD using the qmpy API\cite{kirklin2015open}showed no tie-lines connecting any of them to Si, indicating they are unstable when in contact with Si. This is consistent with the published results \cite{atanassova1998x, schlom2002thermodynamic, alers1998intermixing, perego2008energy, mccurdy2004investigation, george2013preferentially, hu2003interface, goncharova2006interface}. We also analyzed Gd$_2$O$_3$, a high $\epsilon$ ($\sim$ 20\cite{zhou2004properties}) that is proven to be stable against Si\cite{kwo2001properties}, and found that a tie-line does exist between Si and Gd$_2$O$_3$. These phase diagram plots are provided in Supplementary Figure 6. In Figure \ref{fig:phase_diagram}, we report a phase diagram to assess the stability of newly discovered high-$\epsilon$ dielectrics - HoClO and Eu$_5$SiCl$_6$O$_4$. The phase diagram shows that both these materials are thermodynamically stable with the semiconductors such as Si, Ge, GaAs, GaN, and SiC at 0K, a requirement for them to be used in microelectronic devices where an interface with one of the common semiconductors is often necessary\cite{robertson2005high}. The next most promising candidate, tetragonal Tl$_3$PbBr$_5$, has a very large $\epsilon$ (101) but possesses a relatively smaller band gap (2.9 eV) and is computed to be thermodynamically metastable at 0K ($E_{\text{hd}}$ = 16 meV per atom) according to the data obtained from the OQMD. Tl$_3$PbBr$_5$ is also reported in literature to have been experimentally synthesized\cite{keller1981darstellung, denysyuk2013electronic, ferrier2006tl3pbbr5}, without any mention of its dielectric properties.
    
\section{Discussion}
    We report the identification of three dielectric materials that contain a combination of high dielectric constant and large band gap - HoClO($\epsilon$=75, $E_{\text{g}}$=5.2 eV), Eu$_5$SiCl$_6$O$_4$($\epsilon$=69, $E_{\text{g}}$=5.5 eV), and Tl$_3$PbBr$_5$($\epsilon$=101, $E_{\text{g}}$=2.9 eV). These compounds modify the Pareto-front of previously known high-throughput dielectric constants data available from the MP database. Our screening strategy also uncovers four other dielectric materials with large $E_{\text{g}}$ and moderately large $\epsilon$ - Sr$_2$LuBiO$_6$($\epsilon$=24, $E_{\text{g}}$=2.4 eV), Bi$_5$IO$_7$($\epsilon$=36, $E_{\text{g}}$=2.7 eV), Bi$_3$ClO$_4$($\epsilon$=39, $E_{\text{g}}$=2.3 eV), and Bi$_3$BrO$_4$($\epsilon$=39, $E_{\text{g}}$=2.3 eV) - at the cost of conducting only 17 DFPT calculations overall. We utilize the data available in the  open-source databases (OQMD, MP) to build a statistical optimization model and use it to select the best candidates after searching among 11,102 stable non-metals that are available in the OQMD. Among the newly discovered dielectrics, two mixed-anionic materials - HoClO and Eu$_5$SiCl$_6$O$_4$ are shown to have tie-lines with multiple, commonly used semiconductors on their phase diagrams, that indicate their thermodynamic equilibrium. 
    
    The presence of rare-earth elements such as Ho and Eu in dielectrics can be a challenge for their use in practical applications. However, the ongoing efforts toward increasing their availability such as efficient recycling of rare-earth materials\cite{qiu2019economic,amato2019sustainability} can result in a sufficient supply of elements for mass production of small electronic components. In particular, Ho is an underutilized element in the industry\cite{thornton2015homely} even though it is more abundant in the earth's crust than other widely mined elements such as Mo, Bi, and precious metals\cite{yaroshevsky2006abundances}. Eu is more abundant on earth's crust than Ho and some of the  heavily mined elements such as W and As\cite{yaroshevsky2006abundances}. Hence, an active exploration of cheaper and easier extraction methods for rare earth elements may make it feasible to include them in mass-produced electronics in the near future. The presence of toxic elements such as Pb and Tl can stand as a barrier against including Tl$_3$PbBr$_5$ in consumer electronics. 
    Since mixed-anionic materials are an emerging class of functional materials, our identification of promising dielectric materials in this family opens up further research opportunities on rational design of high-performance dielectrics and their experimental characterizations.
    
    We also assessed the thermodynamic stability  of the new dielectrics by creating a large convex hull diagram containing the best two new dielectrics (HoClO and Eu$_5$SiCl$_6$O$_4$) and several commonly used materials in electronics. The relevance of this analysis is also provided in detail along with examples of previously reported high-$\epsilon$ dielectrics\cite{atanassova1998x, schlom2002thermodynamic, alers1998intermixing, perego2008energy, mccurdy2004investigation, george2013preferentially, hu2003interface, goncharova2006interface} that were later found out to be unstable when in contact with common electronic component materials such as SiO$_2$. Our convex hull analysis indicates that both HoClO and Eu$_5$SiCl$_6$O$_4$ are stable against the common electronic materials that we  considered.
    
    To understand what features of HoClO, Eu$_5$SiCl$_6$O$_4$, and Tl$_3$PbBr$_5$ make them the best dielectric candidates in this study, we have  calculated their electronic structures and partial density of states (Supplementary Figure 5). Our analysis shows that the top of the valence bands and bottom of the conduction bands in these compounds consists of primarily the contributions from the anions (Cl, Br) and cations (Ho, Eu, Tl), respectively. This analysis indicates that having lighter anions (such as Cl, Br) is advantageous as their valence orbitals making up the valence band edge in those compounds will have lower energies, hence, a relatively larger band gap that is desired in high-$\epsilon$ materials.
    
    In addition to the identification of high-dielectrics, we successfully  demonstrated an implementation of cross-database statistical design for computational materials selection. Datasets from the MP and OQMD repositories are used in this work as training data and search-space, respectively. The successful identification of new materials from such a workflow is a another motivation for actively moving toward the interoperability of materials databases, which is one of the four pillars of FAIR data principles\cite{wilkinson2016fair} in the scientific data management. Therefore,  better interoperability across databases amplifies the flexibility in utilizing materials data while solving a complex materials problem.
    
    Lastly, this work also stands as an example of practical implementation of a computational design strategy for property optimization via data-informed material selection. A multi-objective optimization problem (maximizing $\epsilon$ and $E_{\text{g}}$) is converted into  a single objective optimization using statistical methods (maximizing $\epsilon$) combined with explicit constraining of band gap values (higher $E_{\text{g}}$) among materials since $E_{\text{g}}$ is already available for all materials in the search-space. The deviation  from the ideal, statistically benchmarked multi-objective optimization workflows\cite{gopakumar2018multi} enabled the efficient utilization of resources and resulted in the identification of three high-$\epsilon$ dielectrics at the cost of just 17 new DFPT calculations.

\section{Methods}

\subsection{ANN modeling}
    The individual models in the ANN-ensemble consisted of a single hidden layer with the number of neurons in the range of $10^2$. The exact number of neurons varied randomly within a small range (10-30) to avoid any bias that may arise from model architecture since the subset of training-data for each ANN was randomly sampled. Each ANN-ensemble consisted of 2000 independent ANNs. Thus, the $\epsilon$-distribution for each material consisted of 2000 independent $\epsilon$ predictions. A new ANN-ensemble was created and trained for each new design cycle to learn the incremented training-data. The Nadam optimizer is used for network optimization during the training. Both L2 layer regularization and early-stopping callback as implemented in Keras\cite{chollet2015keras}, are implemented for each ANN in the ensemble to prevent the over-fitting. On average, it took between 300 to 400 epochs to reach the local minimum of the loss function. Each epoch is a full iteration of fitting the training-data to update the internal weights of an ANN. Validation details of one of the randomly chosen ANN models from the ensemble are plotted in Figure \ref{fig:model_ei}a for reference. Feature dimensional reduction prior to the training of ANNs was done using principal component analysis algorithm implemented in scikit-learn\cite{scikit-learn}. Model validation during the training of one of the 2000 ANN models in the second design cycle is plotted in Figure \ref{fig:model_ei}a. 

\subsection{DFPT calculations}
    We performed all DFT calculations calculations using the Vienna Ab-initio Simulation Package (VASP)\cite{kresse1996efficiency, kresse1996efficient} with potentials derived using the projector-augmented wave (PAW)\cite{kresse1999ultrasoft, blochl1994projector} method. We calculated the total dielectric constant (sum of electronic and ionic components) values for selected materials using DFPT as implemented in VASP. All the compounds were fully relaxed before the dielectric calculations.  We used an energy cut-off of 520 eV, k-mesh of 6000 k-points per reciprocal atom (KPPRA),  and an  energy-threshold of $ 10^{-8}$ eV during the self-consistent calculations. The forces on the atoms after structural relaxations were less than 10$^{-3}$ eV \AA$^{-1}$. We used the generalized gradient approximation\cite{perdew1996generalized} (GGA) to approximate the exchange-correlation energies of the electrons. A detailed discussion on DFPT calculations is provided in the Supplementary Methods section included within the Supplementary Material. We did DFPT calculations on a set of well-known dielectrics and a few rare-earth compounds, and benchmarked the results against previously reported results in literature. These results indicate the reliability of our calculated $\epsilon$ values, which are provided in Supplementary Table 2. Specifically, two rare-earth oxides (EuO and Ho$_2$O$_3$) and one rare-earth halide (EuF$_2$) were benchmarked to test the accuracy of the standard DFPT calculations in modelling these compounds. Furthermore, our calculations reveal that no imaginary phonon modes appear in HoClO, Eu$_5$SiCl$_6$O$_4$, and Tl$_3$PbBr$_5$, the best high-$\epsilon$ materials identified in this work. More details are provided in Supplementary Table 1 and Supplementary Figure 4.

\section{Data Availability}
    The data used in building statistical design models are open-sourced and available via OQMD and Materials Project databases.
    
    Other data that support the findings of this study are available from the corresponding author upon reasonable request.
    
    

\section{Acknowledgements}
    This work was funded by the SAMSUNG Global Research Outreach Program, and the U.S. Department of Commerce, National Institute of Standards and Technology as part of the Center for Hierarchical Materials Design (CHiMaD) award 70NANB14H012. 
    We acknowledge the computing resources provided by (1) the National Energy Research Scientific Computing Center (NERSC), a U.S. Department of Energy Office of Science User Facility operated under Contract No. DE-AC02-05CH11231, (2) Quest high-performance computing facility at Northwestern University which is jointly supported by the Office of the Provost, the Office for Research, and Northwestern University Information Technology, and (3) the Extreme Science and Engineering Discovery Environment (National Science Foundation Contract ACI-1548562)

\section{Author contributions}
    A.M.G. devised computational strategies, wrote the manuscript and conducted the calculations. KP provided important hands-on guidance in calculations and theoretical understanding. A.M.G. and C.W modeled the project and analyzed the results. All authors have reviewed the manuscript. 

\section{Competing Interests}
    The authors declare no competing financial or non-financial interests.


\newpage
\section{Table Titles and Captions}

\begin{table}[h]
\begin{center}
\begin{minipage}{\textwidth}
\caption{The Pareto-front of dielectric materials dataset from Materials Project}\label{tab:fullMPpf}%
    \begin{tabular}{|llll|}
    \hline
    MP ID   & Material  & $E_g$ (eV) & \textbf{$\epsilon$} \\ \hline \hline
    mp-1138  & LiF       & 8.7 & 9.3   \\ 
    mp-13948 & Cs$_2$HfF$_6$   & 7.2 & 9.3   \\ 
    mp-13947 & Rb$_2$HfF$_6$   & 7.1 & 9.4   \\ 
    mp-7104  & CsCaF$_3$    & 6.9 & 9.7   \\ 
    mp-5347  & KAlF$_4$     & 6.8 & 10.6    \\ 
    mp-10250 & BaLiF$_3$    & 6.6 & 14.8  \\ 
    mp-3654  & RbCaF$_3$    & 6.4 & 18.9  \\ 
    mp-8455  & CsF       & 5.9 & 20.3  \\ 
    mp-28243 & RbLiCl$_2$   & 5.1 & 54.5  \\ 
    mp-5606  & AlTlF$_4$    & 4.2 & 96.9    \\ 
    mp-23092 & Ba$_2$TaBiO$_6$ & 2.6 & 99.9  \\ 
    mp-27832 & Tl$_2$SnCl$_6$  & 2.5 & 100.8    \\ 
    mp-3614  & KTaO$_3$     & 2.1 & 639.9 \\ 
    mp-2657  & TiO$_2$      & 1.8  & 988.0 \\ 
    \hline
    \end{tabular}
\footnotetext{MP ID corresponds to the unique ID of material in the repository, $E_{\text{g}}$ is the band gap  energy, and $\epsilon$ corresponds to the largest eigenvalue in the dielectric constant tensor}
\end{minipage}
\end{center}
\end{table}

\begin{table}[h]
\begin{center}
\begin{minipage}{\textwidth}
\caption{Dielectric constants of 17 materials calculated using DFT in this work}\label{tab:all_new_meas}%
    \begin{tabular}{|lllllll|}
    \hline
    \multicolumn{1}{|p{1.2cm}}{\centering OQMD \\ ID} 
    & Material 
    & \multicolumn{1}{p{0.7cm}}{\centering $E_g$ \\ (eV)}
    & \textbf{$\epsilon_x$} & \textbf{$\epsilon_y$} & \textbf{$\epsilon_z$} & Cycle \\ \hline \hline
    681780  & CaVO$_3$      & 0.4 & 4.7   & 4.5   & 4.5   & 1 \\
    14476   & Sr$_2$VN$_3$     & 1.8 & 28.8  & 16.5  & 16.0  & 1 \\
    13450   & BaZrN$_2$     & 1.2 & 31.2  & 31.2  & 21.7  & 1 \\
    1104204 & HoN        & 0.4 & 376.9 & 373.0 & 372.7 & 1 \\
    649584  & Bi$_2$SeO$_2$    & 0.5 & 377.3 & 371.8 & 118.2 & 1 \\ \hline
    19571   & \textbf{Sr$_2$LuBiO$_6$}  & 2.4 & 24.1  & 19.4  & 18.7  & 2 \\
    5958    & \textbf{Bi$_5$IO$_7$}     & 2.7 & 35.8  & 28.2  & 23.1  & 2 \\
    24994   & \textbf{Bi$_3$ClO$_4$}    & 2.3 & 38.9  & 24.2  & 25.7  & 2 \\
    22697   & \textbf{Bi$_3$BrO$_4$}    & 2.3 & 39.0  & 23.7  & 22.1  & 2 \\
    118234  & \textbf{Tl$_3$PbBr$_5$}   & 2.9 & 100.8 & 36.4  & 36.4  & 2 \\ \hline
    11916   & Eu$_4$Cl$_6$O    & 5.3 & 7.4   & 7.3   & 5.5   & 3 \\
    18953   & EuClF      & 5.6 & 11.1  & 11.1  & 10.4  & 3 \\
    646321  & Rb$_2$PrCl$_5$   & 5.1 & 12.2  & 11.0  & 8.9   & 3 \\
    15191   & Cs$_2$NaCeCl$_6$ & 5.1 & 13.2  & 13.2  & 13.2  & 3 \\
    4063    & EuCl$_2$      & 5.2 & 15.6  & 12.9  & 11.8  & 3 \\
    24611   & \textbf{Eu$_5$SiCl$_6$O$_4$} & 5.5 & 69.3  & 15.1  & 12.9  & 3 \\
    13689   & \textbf{HoClO}      & 5.2 & 75.1  & 37.9  & 15.2  & 3 \\ \hline
    \end{tabular}
\footnotetext{OQMD ID refers to the materials' unique entry ID in the OQMD database, $E_{\text{g}}$ refers to the band gap energy in eV, $\epsilon_{x,y,z}$ refers to the three eigenvalues (xx, yy, zz) of the  of dielectric constant tensor, and the Cycle mentions the design cycle when the material was selected for the calculations of dielectric constant using DFPT. The values $\epsilon_{x,y,z}$ are ordered in such a way that $\epsilon_x > \epsilon_y > \epsilon_z$. The best seven materials found in this work are highlighted in bold letters.}
\end{minipage}
\end{center}
\end{table}

\newpage
\section{Figure Captions}

\begin{figure*}[!htp]
    \centering
    \includegraphics[width=0.9\linewidth]{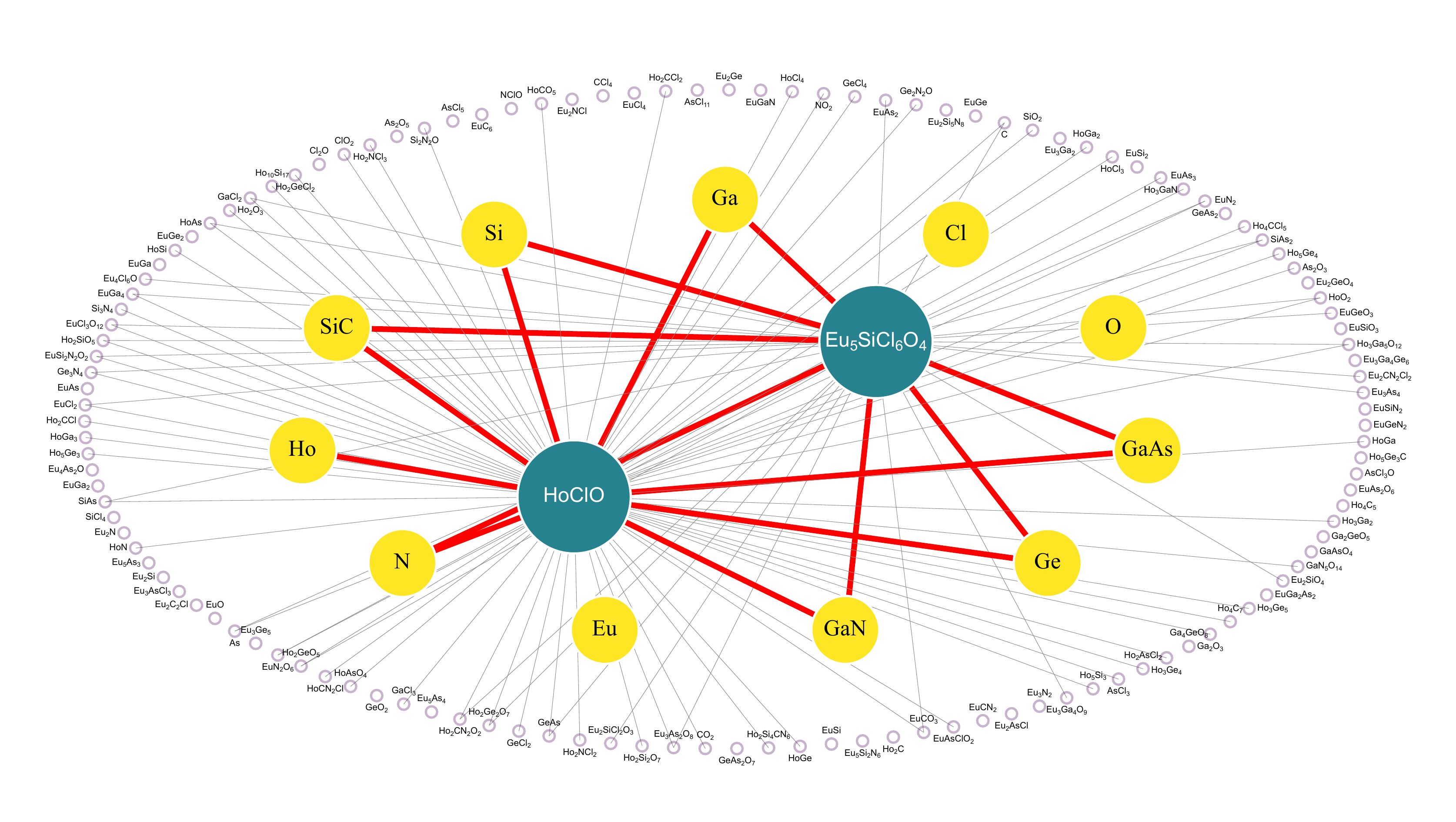}
    \caption{\textbf{Phase diagram of all stable compounds in Ho-Cl-O-Eu-Si-Ge-Ga-As-C-N phase-space from OQMD (as of January, 2022)}. The two new most promising dielectrics, HoClO and Eu$_5$SiCl$_6$O$_4$ are plotted in large green circles in the center. The elements (Ho, Eu, Si, Cl, Ge, Ga, As, C, N and O) and semiconductors of interest (Si, Ge, GaAs, SiC, and GaN) are plotted in the middle layer in medium-sized yellow circles. All other stable compounds in the phase diagram are plotted in small dark circles in the outermost layer. Tie-lines between the new dielectrics and the semiconductors or elements are shown as thick red lines. Other tie-lines from the dielectrics to rest of the stable materials in outer layer are drawn as narrow gray lines. Another 2326 tie-lines exist in this phase-diagram that do not include either of the dielectrics, and are not shown in this network-plot for better visibility of relevant information. The elements and compounds without any visible tie-lines in the outermost layer do not have tie-lines with HoClO or Eu$_5$SiCl$_6$O$_4$, but they have tie-lines with some of the other materials in the outer layer - making them part of this phase diagram. There exist a tie-line from each dielectric material to each semiconductor that is considered here for comparison. This indicates that the both HoClO and Eu$_5$SiCl$_6$O$_4$  are in thermodynamic equilibrium with Si, Ge, GaAs, GaN, and SiC at 0K. The thermodynamic stability is a requirement for dielectrics that needs to form a stable interface with semiconductors in electronic applications\cite{robertson2005high}. 
    }
    \label{fig:phase_diagram}
\end{figure*}

\end{document}